\documentclass[10pt,a4paper,onecolumn]{article}
\usepackage{marginnote}
\usepackage{graphicx}
\usepackage{xcolor}
\usepackage{authblk,etoolbox}
\usepackage{titlesec}
\usepackage{calc}
\usepackage{tikz}
\usepackage{hyperref}
\hypersetup{colorlinks,breaklinks=true,
            urlcolor=[rgb]{0.0, 0.5, 1.0},
            linkcolor=[rgb]{0.0, 0.5, 1.0}}
\usepackage{caption}
\usepackage{tcolorbox}
\usepackage{amssymb,amsmath}
\usepackage{ifxetex,ifluatex}
\usepackage{seqsplit}
\usepackage{xstring}

\usepackage{float}
\let\origfigure\figure
\let\endorigfigure\endfigure
\renewenvironment{figure}[1][2] {
    \expandafter\origfigure\expandafter[H]
} {
    \endorigfigure
}

\usepackage{fixltx2e} % provides \textsubscript
\usepackage[
  backend=biber,
]{biblatex}
\bibliography{main.bib}

\let\textttOrig=\texttt
\def\texttt#1{\expandafter\textttOrig{\seqsplit{#1}}}
\renewcommand{\seqinsert}{\ifmmode
  \allowbreak
  \else\penalty6000\hspace{0pt plus 0.02em}\fi}

\makeatletter
\let\href@Orig=\href
\def\href@Urllike#1#2{\href@Orig{#1}{\begingroup
    \def\Url@String{#2}\Url@FormatString
    \endgroup}}
\def\href@Notdoi#1#2{\def\tempa{#1}\def\tempb{#2}%
  \ifx\tempa\tempb\relax\href@Urllike{#1}{#2}\else
  \href@Orig{#1}{#2}\fi}
\def\href#1#2{%
  \IfBeginWith{#1}{https://doi.org}%
  {\href@Urllike{#1}{#2}}{\href@Notdoi{#1}{#2}}}
\makeatother

\newlength{\cslhangindent}
\newlength{\csllabelwidth}
\newenvironment{CSLReferences}[3] % #1 hanging-ident, #2 entry spacing
 {% don't indent paragraphs
  \setlength{\parindent}{0pt}
  \ifodd #1 \everypar{\setlength{\hangindent}{\cslhangindent}}\ignorespaces\fi
  \ifnum #2 > 0
  \setlength{\parskip}{#2\baselineskip}
  \fi
 }%
 {}
\usepackage{calc}

\usepackage[top=3.5cm, bottom=3cm, right=1.5cm, left=1.0cm,
            headheight=2.2cm, reversemp, includemp, marginparwidth=4.5cm]{geometry}

\titleformat{\section}
  {\normalfont\sffamily\Large\bfseries}
  {}{0pt}{}
\titleformat{\subsection}
  {\normalfont\sffamily\large\bfseries}
  {}{0pt}{}
\titleformat{\subsubsection}
  {\normalfont\sffamily\bfseries}
  {}{0pt}{}
\titleformat*{\paragraph}
  {\sffamily\normalsize}

\usepackage{fancyhdr}
\makeatletter
\let\ps@plain\ps@fancy
\fancyheadoffset[L]{4.5cm}
\fancyfootoffset[L]{4.5cm}

\definecolor{linky}{rgb}{0.0, 0.5, 1.0}

\newtcolorbox{repobox}
   {colback=red, colframe=red!75!black,
     boxrule=0.5pt, arc=2pt, left=6pt, right=6pt, top=3pt, bottom=3pt}

\newcommand{\ExternalLink}{%
   \tikz[x=1.2ex, y=1.2ex, baseline=-0.05ex]{%
       \begin{scope}[x=1ex, y=1ex]
           \clip (-0.1,-0.1)
               --++ (-0, 1.2)
               --++ (0.6, 0)
               --++ (0, -0.6)
               --++ (0.6, 0)
               --++ (0, -1);
           \path[draw,
               line width = 0.5,
               rounded corners=0.5]
               (0,0) rectangle (1,1);
       \end{scope}
       \path[draw, line width = 0.5] (0.5, 0.5)
           -- (1, 1);
       \path[draw, line width = 0.5] (0.6, 1)
           -- (1, 1) -- (1, 0.6);
       }
   }

\patchcmd{\@maketitle}{center}{flushleft}{}{}
\patchcmd{\@maketitle}{center}{flushleft}{}{}
\patchcmd{\@maketitle}{\LARGE}{\LARGE\sffamily}{}{}
\def\maketitle{{%
  
  \AB@maketitle}}
\makeatletter
\renewcommand\AB@affilsepx{ \protect\Affilfont}
\renewcommand\AB@affilnote[1]{{\bfseries #1}\hspace{3pt}}
\renewcommand{\affil}[2][]%
   {\newaffiltrue\let\AB@blk@and\AB@pand
      \if\relax#1\relax\def\AB@note{\AB@thenote}\else\def\AB@note{#1}%
        \setcounter{Maxaffil}{0}\fi
        \begingroup
        \let\href=\href@Orig
        \let\texttt=\textttOrig
        \let\protect\@unexpandable@protect
        \def\thanks{\protect\thanks}\def\footnote{\protect\footnote}%
        \@temptokena=\expandafter{\AB@authors}%
        {\def\\{\protect\\\protect\Affilfont}\xdef\AB@temp{#2}}%
         \xdef\AB@authors{\the\@temptokena\AB@las\AB@au@str
         \protect\\[\affilsep]\protect\Affilfont\AB@temp}%
         \gdef\AB@las{}\gdef\AB@au@str{}%
        {\def\\{, \ignorespaces}\xdef\AB@temp{#2}}%
        \@temptokena=\expandafter{\AB@affillist}%
        \xdef\AB@affillist{\the\@temptokena \AB@affilsep
          \AB@affilnote{\AB@note}\protect\Affilfont\AB@temp}%
      \endgroup
       \let\AB@affilsep\AB@affilsepx
}
\makeatother

\renewcommand\Affilfont{\sffamily\small\mdseries}
\ifnum 0\ifxetex 1\fi\ifluatex 1\fi=0 % if pdftex
  \usepackage[T1]{fontenc}
  \usepackage[utf8]{inputenc}

\else % if luatex or xelatex
  \ifxetex
    \usepackage{mathspec}
    \usepackage{fontspec}

  \else
    \usepackage{fontspec}
  \fi
  \defaultfontfeatures{Ligatures=TeX,Scale=MatchLowercase}

\fi
\IfFileExists{upquote.sty}{\usepackage{upquote}}{}
\IfFileExists{microtype.sty}{%
\usepackage{microtype}
\UseMicrotypeSet[protrusion]{basicmath} % disable protrusion for tt fonts
}{}

\usepackage{hyperref}
\hypersetup{unicode=true,
            pdftitle={denmarf: a Python package for density estimation using masked autoregressive flow},
            pdfborder={0 0 0},
            breaklinks=true}
\let\addcontentslineOrig=\addcontentsline
\def\addcontentsline#1#2#3{\bgroup
  \let\texttt=\textttOrig\addcontentslineOrig{#1}{#2}{#3}\egroup}
\let\markbothOrig\markboth
\def\markboth#1#2{\bgroup
  \let\texttt=\textttOrig\markbothOrig{#1}{#2}\egroup}
\let\markrightOrig\markright
\def\markright#1{\bgroup
  \let\texttt=\textttOrig\markrightOrig{#1}\egroup}

\usepackage{graphicx,grffile}
\makeatletter
\def\maxwidth{\ifdim\Gin@nat@width>\linewidth\linewidth\else\Gin@nat@width\fi}
\def\maxheight{\ifdim\Gin@nat@height>\textheight\textheight\else\Gin@nat@height\fi}
\makeatother
\setkeys{Gin}{width=\maxwidth,height=\maxheight,keepaspectratio}
\IfFileExists{parskip.sty}{%
\usepackage{parskip}
}{% else
\setlength{\parindent}{0pt}
\setlength{\parskip}{6pt plus 2pt minus 1pt}
}
\ifx\paragraph\undefined\else
\let\oldparagraph\paragraph
\renewcommand{\paragraph}[1]{\oldparagraph{#1}\mbox{}}
\fi
\ifx\subparagraph\undefined\else
\let\oldsubparagraph\subparagraph
\renewcommand{\subparagraph}[1]{\oldsubparagraph{#1}\mbox{}}
\fi

\title{denmarf: a Python package for density estimation using masked
autoregressive flow}

        \author[1]{Rico K. L. Lo}
    
      \affil[1]{LIGO Laboratory, California Institute of Technology,
Pasadena, California 91125, USA}
  \date{\vspace{-7ex}}

\begin{document}
\maketitle

\marginpar{

  \begin{flushleft}
  %\hrule
  \sffamily\small

  {\bfseries DOI:} \href{https://doi.org/10.xxxxxx/draft}{\color{linky}{10.xxxxxx/draft}}

  \vspace{2mm}

  {\bfseries Software}
  \begin{itemize}
    \setlength\itemsep{0em}
    \item \href{N/A}{\color{linky}{Review}} \ExternalLink
    \item \href{https://github.com/ricokaloklo/denmarf}{\color{linky}{Repository}} \ExternalLink
    \item \href{N/A}{\color{linky}{Archive}} \ExternalLink
  \end{itemize}

  \vspace{2mm}

  \par\noindent\hrulefill\par

  \vspace{2mm}

  {\bfseries Editor:} \href{N/A}{Pending
Editor} \ExternalLink \\
  \vspace{1mm}
    {\bfseries Reviewers:}
  \begin{itemize}
  \setlength\itemsep{0em}
    \item \href{N/A}{@Pending
Reviewers}
    \end{itemize}
    \vspace{2mm}

  {\bfseries Submitted:} N/A\\
  {\bfseries Published:} N/A

  \vspace{2mm}
  {\bfseries License}\\
  Authors of papers retain copyright and release the work under a Creative Commons Attribution 4.0 International License (\href{http://creativecommons.org/licenses/by/4.0/}{\color{linky}{CC BY 4.0}}).

  \end{flushleft}
}

\hypertarget{summary}{%
\section{Summary}\label{summary}}

Masked autoregressive flow (MAF) (Papamakarios et al., 2017) is a
state-of-the-art non-parametric density estimation technique. It is
based on the idea (known as a normalizing flow) that a simple base
probability distribution can be mapped into a complicated target
distribution that one wishes to approximate, using a sequence of
bijective transformations (Tabak \& Turner, 2013; Tabak \&
Vanden-Eijnden, 2010). The \texttt{denmarf} package provides a
\texttt{scikit-learn}-like interface in Python for researchers to
effortlessly use MAF for density estimation in their applications to
evaluate probability densities of the underlying distribution of a set
of data and generate new samples from the data, on either a CPU or a
GPU, as simple as
\texttt{from\ denmarf\ import\ DensityEstimate;\ de\ =\ DensityEstimate().fit(X)}.
The package also implements logistic transformations to facilitate the
fitting of bounded distributions.

\hypertarget{statement-of-need}{%
\section{Statement of need}\label{statement-of-need}}

There are a number of ways to perform density estimation in a
non-parametric fashion, one of which is kernel density estimation (KDE).
Suppose we have a set of \(D\)-dimensional data of size \(N\),
\(\left( \vec{x}_{1}, \vec{x}_{2}, \dots, \vec{x}_{N} \right)\),
i.e.~\(\vec{x}_{i}\) is a \(D\)-dimensional vector where
\(i \in \left[ 1, N \right]\) that follows the probability distribution
\(f(\vec{x})\) we wish to approximate. The kernel density estimate
\(\hat{f}_{\rm KDE}\) using those input data is given by (Scott, 1992)
\begin{equation}
\label{eq:KDE}
  \hat{f}_{\rm KDE}(\vec{x}) = \dfrac{1}{N} \sum_{i=1}^{N} K(\vec{x} - \vec{x}_{i}),
\end{equation} where \(K\) is the kernel function that depends on the
distance between the evaluation point \(\vec{x}\) and the input data
point \(\vec{x}_{i}\). There are many implementations of KDE in Python,
such as \texttt{scipy.stats.gaussian\_kde} (Virtanen et al., 2020),
\texttt{sklearn.neighbors.KernelDensity} (Pedregosa et al., 2011) and
\texttt{kalepy} (Kelley, 2021). The cost of \(M\) such evaluations using
\autoref{eq:KDE} is therefore \(O(MND)\). This can be slow if we need to
evaluate the KDE of a large data set (i.e.~large \(N\)) many times
(i.e.~large \(M\)). For instance, one might wish to evaluate the
probability density, estimated from a large number (\(N \sim 10^7\)) of
simulated lensed astronomical objects, of two lensed images of a
background object having certain magnifications over a set of possible
(\(M \sim 10^5\)) values.

\begin{figure}
\centering
\includegraphics[width=0.8\textwidth,height=\textheight]{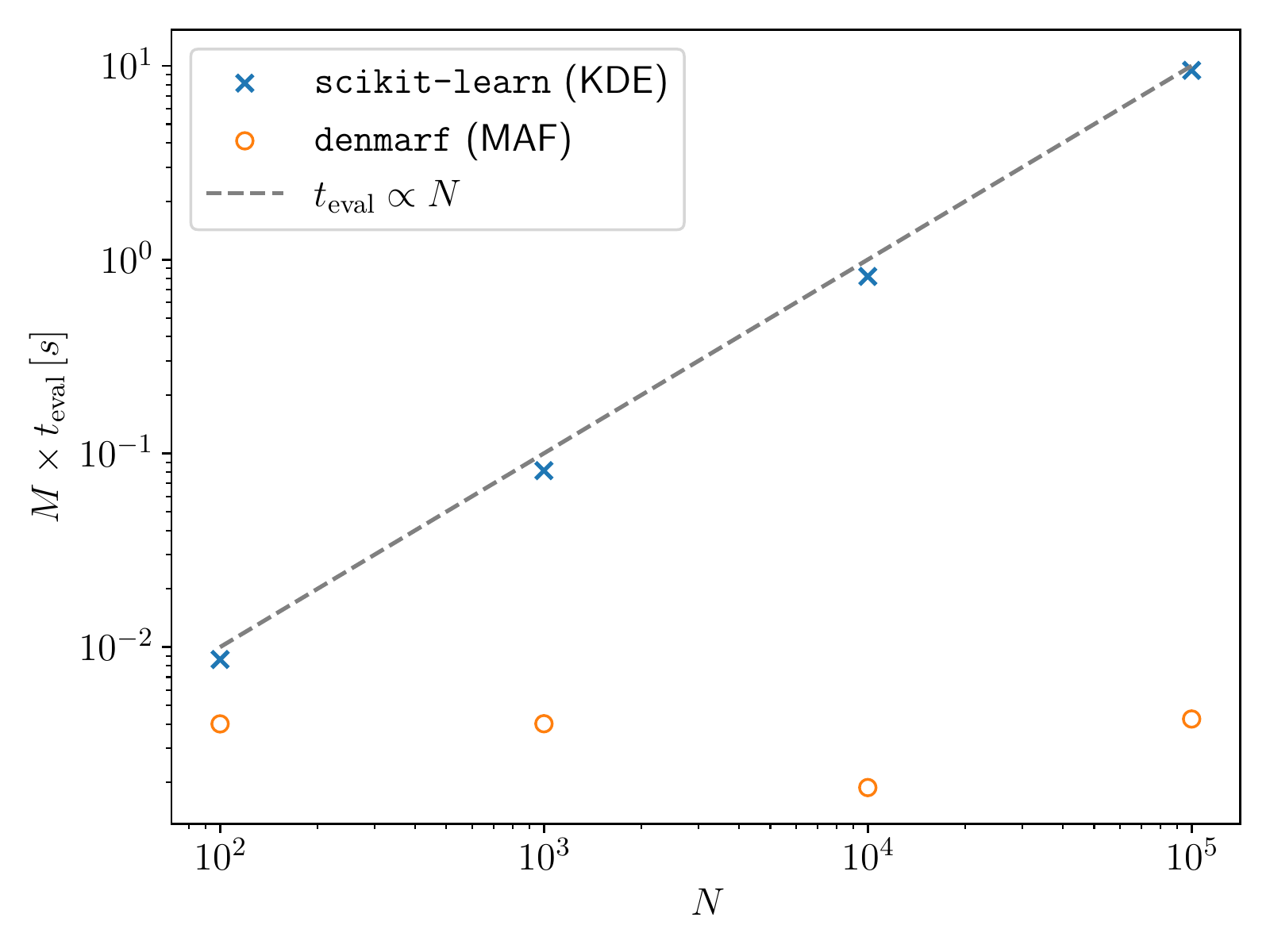}
\caption{Computation cost for \(M = 1000\) evaluations of the density
estimate from data of size \(N\) using KDE with \texttt{scikit-learn}
and that using MAF with \texttt{denmarf} respectively. We can see that
the evaluation cost using KDE scales with \(N\) while that using MAF is
independent of \(N\). \label{fig:timing}}
\end{figure}

However with MAF, an evaluation of the estimated density is independent
of \(N\). Suppose \(T(\vec{x})\) maps the target distribution
\(f(\vec{x})\) into the base distribution \(u\), usually chosen as a
\(D\)-dimensional standard normal distribution, then the density
estimate using MAF \(\hat{f}_{\rm MAF}\) is given by \begin{equation}
  \hat{f}_{\rm MAF}(\vec{x}) = u(T(\vec{x}))|J_{T}(\vec{x})|,
\end{equation} where \(|J_{T}|\) is the Jacobian determinant of the
mapping, and note that there is no summation over the \(N\) input data.
\autoref{fig:timing} shows the computational cost for \(M = 1000\)
evaluations of the density estimate from data of size \(N\) using KDE
and that using MAF respectively. We can see that the evaluation cost
using KDE scales with \(N\) while that using MAF is indeed independent
of \(N\).

While it is relatively straightforward to implement a routine to perform
density estimation using MAF with the help of deep learning libraries
such as \texttt{TensorFlow} (Abadi et al., 2015) and \texttt{PyTorch}
(Paszke et al., 2017), the technical hurdle of leveraging MAF for people
not well-versed in those libraries remains high. The \texttt{denmarf}
package is designed to be an almost drop-in replacement of the
\texttt{sklearn.neighbors.KernelDensity} module to lower the technical
barrier and enable researchers to apply MAF for density estimation
effortlessly.

New samples \(\vec{x}_{i}\) can be generated from the approximated
distribution by first drawing samples \(\vec{y}_{i}\) from the base
distribution \(u(\vec{y})\) and then transforming them with the inverse
mapping \(T^{-1}\), i.e. \begin{equation}
  \vec{x}_{i} = T^{-1}(\vec{y}_{i}).
\end{equation} Indeed, if the transformations are bijective (i.e.~both
surjective and injective) then we can always find \(\vec{x}_{i}\) such
that \(\vec{y}_{i} = T(\vec{x}_{i})\). This could potentially be a
problem for input data \(\vec{x}_{i}\) that are bounded, since in MAF
\(T\) is only rescaling and shifting (i.e.~an affine transformation) and
\(u\) is usually a normal distribution which is unbounded. To solve this
problem, \texttt{denmarf} will logit-transform the input data first if
the underlying distribution should be bounded, and the logit-transformed
data become unbounded. \texttt{denmarf} will automatically include the
extra Jacobian from the logit transformation during density evaluations
and perform inverse logit transformation after sample regenerations.

\hypertarget{acknowledgements}{%
\section{Acknowledgements}\label{acknowledgements}}

The author would like to acknowledge support from National Science
Foundation Awards No.~PHY-1912594 and No.~PHY-2207758. The author is
also grateful for computational resources provided by the LIGO
Laboratory and supported by NSF Grants No.~PHY-0757058 and
No.~PHY-0823459. The current version of \texttt{denmarf} uses
\texttt{pytorch-flows} as the backbone (Kostrikov, 2021).

\hypertarget{references}{%
\section*{References}\label{references}}
\addcontentsline{toc}{section}{References}

\hypertarget{refs}{}
\begin{CSLReferences}{1}{0}
\leavevmode\hypertarget{ref-tensorflow2015-whitepaper}{}%
Abadi, M., Agarwal, A., Barham, P., Brevdo, E., Chen, Z., Citro, C.,
Corrado, G. S., Davis, A., Dean, J., Devin, M., Ghemawat, S.,
Goodfellow, I., Harp, A., Irving, G., Isard, M., Jia, Y., Jozefowicz,
R., Kaiser, L., Kudlur, M., \ldots{} Zheng, X. (2015).
\emph{{TensorFlow}: Large-scale machine learning on heterogeneous
systems}. \url{https://www.tensorflow.org/}

\leavevmode\hypertarget{ref-Kelley2021}{}%
Kelley, L. Z. (2021). Kalepy: A python package for kernel density
estimation, sampling and plotting. \emph{Journal of Open Source
Software}, \emph{6}(57), 2784. \url{https://doi.org/10.21105/joss.02784}

\leavevmode\hypertarget{ref-ikostrikov}{}%
Kostrikov, I. (2021). {pytorch-flows: PyTorch implementations of
algorithms for density estimation}. In \emph{GitHub repository}.
\url{https://github.com/ikostrikov/pytorch-flows}; GitHub.

\leavevmode\hypertarget{ref-NIPS2017_6c1da886}{}%
Papamakarios, G., Pavlakou, T., \& Murray, I. (2017). Masked
autoregressive flow for density estimation. In I. Guyon, U. V. Luxburg,
S. Bengio, H. Wallach, R. Fergus, S. Vishwanathan, \& R. Garnett (Eds.),
\emph{Advances in neural information processing systems} (Vol. 30).
Curran Associates, Inc.
\url{https://proceedings.neurips.cc/paper_files/paper/2017/file/6c1da886822c67822bcf3679d04369fa-Paper.pdf}

\leavevmode\hypertarget{ref-paszke2017automatic}{}%
Paszke, A., Gross, S., Chintala, S., Chanan, G., Yang, E., DeVito, Z.,
Lin, Z., Desmaison, A., Antiga, L., \& Lerer, A. (2017). Automatic
differentiation in PyTorch. \emph{NIPS-w}.

\leavevmode\hypertarget{ref-scikit-learn}{}%
Pedregosa, F., Varoquaux, G., Gramfort, A., Michel, V., Thirion, B.,
Grisel, O., Blondel, M., Prettenhofer, P., Weiss, R., Dubourg, V.,
Vanderplas, J., Passos, A., Cournapeau, D., Brucher, M., Perrot, M., \&
Duchesnay, E. (2011). Scikit-learn: Machine learning in {P}ython.
\emph{Journal of Machine Learning Research}, \emph{12}, 2825--2830.

\leavevmode\hypertarget{ref-scott1992multivariate}{}%
Scott, D. W. (1992). \emph{Multivariate density estimation: Theory, practice, and visualization}. Wiley. ISBN: 9780471547709

\leavevmode\hypertarget{ref-https:ux2fux2fdoi.orgux2f10.1002ux2fcpa.21423}{}%
Tabak, E. G., \& Turner, C. V. (2013). A family of nonparametric density
estimation algorithms. \emph{Communications on Pure and Applied
Mathematics}, \emph{66}(2), 145--164.
\url{https://doi.org/10.1002/cpa.21423}

\leavevmode\hypertarget{ref-cmsux2f1266935020}{}%
Tabak, E. G., \& Vanden-Eijnden, E. (2010). {Density estimation by dual
ascent of the log-likelihood}. \emph{Communications in Mathematical
Sciences}, \emph{8}(1), 217--233.

\leavevmode\hypertarget{ref-2020SciPy-NMeth}{}%
Virtanen, P., Gommers, R., Oliphant, T. E., Haberland, M., Reddy, T.,
Cournapeau, D., Burovski, E., Peterson, P., Weckesser, W., Bright, J.,
van der Walt, S. J., Brett, M., Wilson, J., Millman, K. J., Mayorov, N.,
Nelson, A. R. J., Jones, E., Kern, R., Larson, E., \ldots{} SciPy 1.0
Contributors. (2020). {{SciPy} 1.0: Fundamental Algorithms for
Scientific Computing in Python}. \emph{Nature Methods}, \emph{17},
261--272. \url{https://doi.org/10.1038/s41592-019-0686-2}

\end{CSLReferences}

\end{document}